\documentclass[10pt]{article}

\usepackage{amsmath}
\usepackage{amssymb}

\usepackage{graphicx}

\usepackage{cite}

\usepackage{color} 

\usepackage[labelfont=bf,labelsep=period,justification=raggedright]{caption}

\makeatletter
\renewcommand{\@biblabel}[1]{\quad#1.}
\makeatother

\date{}

\begin{document}

\begin{flushleft}
{\Large
\textbf{A computational model incorporating neural stem cell dynamics reproduces glioma incidence across the
lifespan in the human population}
}
\\
Roman Bauer$^{1}$, 
Marcus Kaiser$^{1,2}$, 
Elizabeth Stoll$^{2\ast}$
\\
\bf{1} Interdisciplinary Computing and Complex BioSystems Research Group (ICOS), School of Computing Science, Newcastle University, Newcastle upon Tyne, Tyne and Wear, United Kingdom
\\
\bf{2} Institute of Neuroscience, Newcastle University, Newcastle upon Tyne, Tyne and Wear, United Kingdom
\\
$\ast$ E-mail: elizabeth.stoll@newcastle.ac.uk
\end{flushleft}

\section*{Abstract}

Glioma is the most common form of primary brain tumor. Demographically, the risk of occurrence
increases until old age. Here we present a novel computational model to reproduce the
probability of glioma incidence across the lifespan. Previous mathematical models explaining
glioma incidence are framed in a rather abstract way, and do not directly relate to empirical
findings. To decrease this gap between theory and experimental observations, we incorporate
recent data on cellular and molecular factors underlying gliomagenesis. Since evidence implicates
the adult neural stem cell as the likely cell-of-origin of glioma, we have incorporated
empirically-determined estimates of neural stem cell number, cell division rate, mutation rate
and oncogenic potential into our model. We demonstrate that our model yields results which match
actual demographic data in the human population. In particular, this model accounts for the
observed peak incidence of glioma at approximately 80 years of age, without the need to assert
differential susceptibility throughout the population. Overall, our model supports the hypothesis
that glioma is caused by randomly-occurring oncogenic mutations within the neural stem cell
population. Based on this model, we assess the influence of the (experimentally indicated) decrease
in the number of neural stem cells and increase of cell division rate during aging. Our model provides
multiple testable predictions, and suggests that different temporal sequences of oncogenic mutations can lead to
tumorigenesis. Finally, we conclude that four or five oncogenic mutations are sufficient for
the formation of glioma.

\section*{Introduction}

Glioma is the most common form of primary brain tumor \cite{Doleceketal2012cbtrus}.
Glioma commonly manifests itself
as a high-grade tumor called glioblastoma, a highly malignant and invasive tumor with
median patient survival of 12 months from diagnosis; lower-grade gliomas increase in
malignancy over time, with associated increases in mortality \cite{OhgakiKleihues2005population}. 

The cellular mechanisms giving rise to glioma are subject to intense research. The
incidence of glioma is not significantly affected by environmental factors such as UV
light and carcinogen exposure, due to the protective influence of the thick skull and
the blood-brain barrier. In addition, there are no known heritable factors in the risk
of glioma occurrence. These tumors appear to arise idiopathically in a random manner
throughout the population \cite{OstromBarnholtzSloan2011current}. Hence, glioma
formation is an ideal test-case for investigating
how fundamental mechanisms on the single-cell level give rise to cancer.

Increasing age is strongly associated with higher incidence and increased malignant
grade for all grades and types of glioma \cite{Porteretal2010prevalence,Barkeretal2001age}.
Age is in fact the single most robust
factor influencing glioma incidence, malignancy, and patient survival
\cite{Doleceketal2012cbtrus,OhgakiKleihues2005population,Porteretal2010prevalence}. Insights into changes
that occur in the aging brain and the cells that originate the tumor are therefore essential for understanding
this increased risk of oncogenic transformation and tumorigenesis.

The putative cell-of-origin of glioma is the neural stem cell (NSC), which normally gives
rise to new neurons and glial cells in the adult brain. Experimentally causing oncogenic mutations in
this lineage leads to the formation of malignant tumors
\cite{Hollandetal2000combined,Alcantaraetal2009malignant,Wangetal2009expression},
and gliomas cluster near
germinal centers of the brain \cite{Limetal2007relationship}. Proliferative cells
within the tumor share immunomarkers
with NSCs \cite{Yuanetal2004isolation,Stilesetal2008glioma}.
NSCs already exist in a proliferative state, are capable of differentiating
into glial cell types, and can migrate through tissue \cite{Stolletal2011aging,Stolletal2011increased}.
Transplantation
of oncogenically-transformed mouse neural stem cells into syngeneic mice reliably leads to
the formation of a tumor which recapitulates the proliferative and invasive phenotype of
human glioma \cite{Mikheevetal2009syngeneic,Mikheevetal2012increased}.
Together, these studies strongly implicate the neural stem cell as
the most likely cell-of-origin of glioma. In this report we show that modeling the accumulation
of random mutations during cell division in this stem cell population can predict glioma
incidence across the lifespan in the human population. In particular, we propose 
a model that accounts for differential weightage and temporal ordering of 
oncogenic mutations.

\section*{Materials and Methods}

The model includes empirical data collected through literature review. The mutation frequency
was taken directly from a published estimate and assumed to be constant across the lifespan
\cite{FrankNowak2004problems}. A small subset of mutations were deemed to have
oncogenic potential in this cellular
compartment while all other mutations are assumed to be neutral for this cancer type
\cite{Bamfordetal2004cosmic}.
In this approach, we used the Poisson-approximation of a binomial distribution
for computing the probabilities to have
x oncogenic mutations. First we
compute the expected number of genetic mutations a cell has had at a certain age, and based
on that then compute the probability of having x oncogenic mutations. The mutation rate is therefore
independent of whether or not the gene is oncogenic.

The exponentially decreasing number of neural stem cells was calculated across the lifespan
based on the published data for human tissue \cite{Sanaietal2011corridors}.
Results of electron microscopy-based
characterization is shown in Figure 3 of \cite{Sanaietal2011corridors},
which used 200 micron thick sections.
Results of immunohistochemistry-based characterization are shown in Figure 1 of \cite{Sanaietal2011corridors},
which used 30 micron thick sections. These data are in agreement - $\sim$144 cells per
200 micron-thick section (averaging the two locations described above) and $\sim$22
DCX+ cells per
$mm^2$ in a 30 micron-thick section (estimated from the graph in Figure 1r of \cite{Sanaietal2011corridors}). These data both yield
approximately 720 DCX+ cells per $mm^3$. To estimate KI67+ proliferative cells, not DCX+ cells,
we multiplied the values for KI67+ cells from the relevant graph (Figure 1s of \cite{Sanaietal2011corridors})  by 33, just as we
multiplied the values for the DCX+ cells from the other graph (Figure 1r of \cite{Sanaietal2011corridors}) by 33. This provides
values  per $mm^3$.  In agreement with the data presented in Figures 1 and 3 of \cite{Sanaietal2011corridors}, their Figure 2c shows
that the tract is 1mm x 1mm wide. It is also 10 mm long (the scale bar represents 500 microns).
So the number of KI67+ cells per $mm^3$ is multiplied again by 10 to estimate the total
number of KI67+ cells. The graph of KI67+ cells at each time point was then extrapolated to
estimate this population across the entire lifespan. Overall, we computed the number of NSCs
at birth to be $237\,600$, which was used as the initial value of the modeled number of
NSCs during aging ($N_{0}$).

\begin{figure}[!ht]
\centering
  \includegraphics[width=1.0\textwidth]{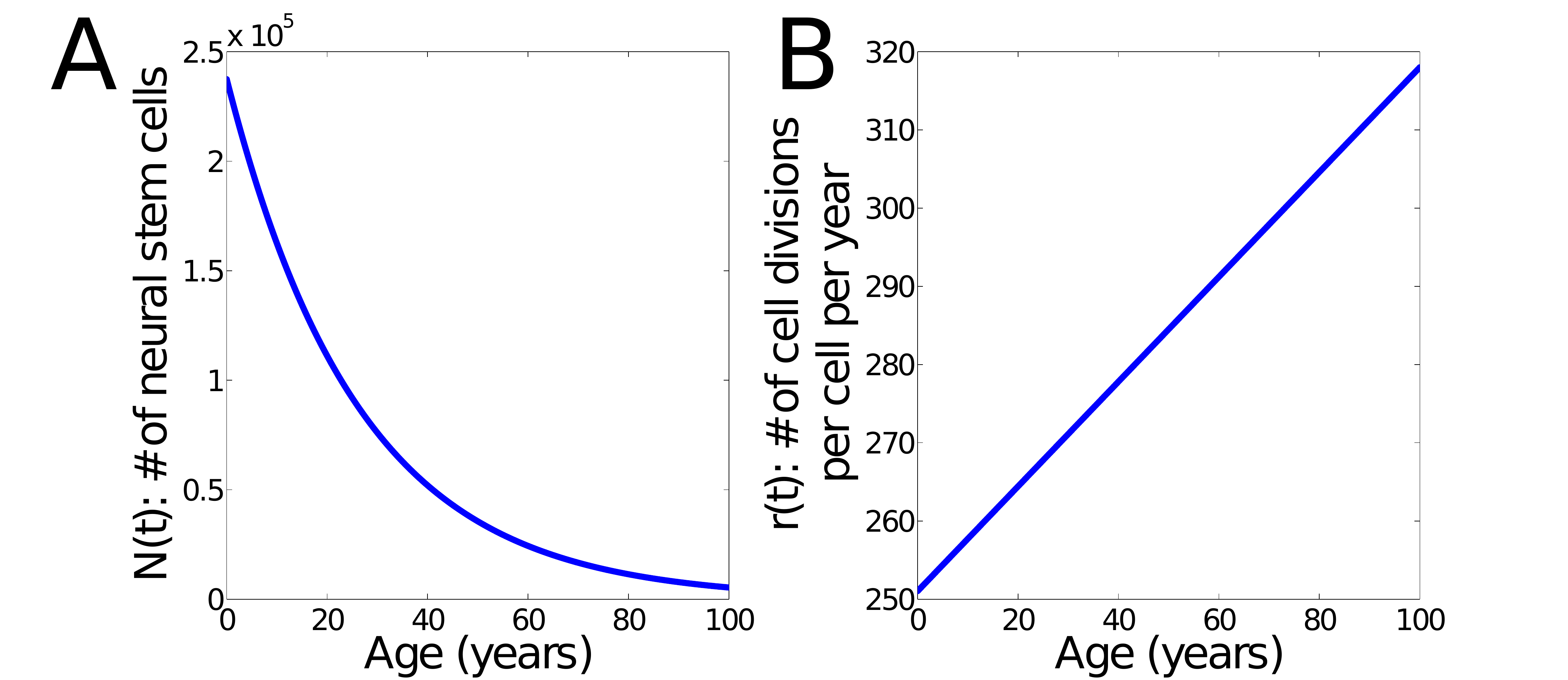}
\caption[mitoticrate_time]{ {\bf Modeled number and cell division rate of NSCs.}
{\bf (A)} Number of NSCs during aging.
The initial number of cells was estimated based on \cite{Sanaietal2011corridors}.
The number of NSCs is given by $N(t)=N_{0}e^{-d\cdot t}$ using $d=0.038$. 
{\bf (B)} Modeled cell division rate over time. As shown in 
\cite{Stolletal2011increased}, NSCs increase their rate
during aging. We have approximated this behavior using a linear 
interpolation from 251 to 318 divisions per cell and year.
\label{fig:mitotic_rate}}
\end{figure}

The cell division rate was calculated in NSCs derived from
the young adult and aged adult mouse brain \cite{Stolletal2011increased}. The
number of cell divisions in a given time was calculated from live-cell time-lapse
imaging over a 48 hour period. Actively-cycling young adult NSCs divided
1.37 times in 48 hours while actively-cycling aged adult NSCs divided 1.74 times
in 48 hours. Adjusted for time, actively-cycling young adult NSCs divide 251
times per year while actively-cycling aged adult NSCs divide 318 times per year. 
For the estimate that is incorporated in the model, we have used a linear interpolation between
these two numbers across the human lifespan.
These estimates were assumed relevant for the population of NSCs in the adult human brain (Fig. \ref{fig:mitotic_rate}).

The model was implemented in MATLAB (Mathworks Inc.). A time step dt of 0.001 years was
used for calculating the prevalence.
The computation of the incidence
was done by computing the numerical differential of the prevalence over time. 
Bootstrapping was used to compute the 95 $\%$ confidence interval of the 
incidence, as shown in Fig. \ref{fig:CI_incidence}. 1000 bootstrap samples 
of size $100\,000$ were computed. 

Two of the model parameters ($d$ describing exponential decrease of NSCs with time
and $s$ included in Eq. \ref{eq:minimum_onecellcancerogenic})
were not assessed from experimental findings. Depending on $r(t)$ and $k_{min}$, different
incidence curves are obtained (i.e. the absolute values and the position of the curve
peak were different). We have adapted $s$ and
$d$ for the different scenarios, in order for the incidence curve
to match with the demographic data \cite{Doleceketal2012cbtrus}. A match could 
only be obtained for $k_{min}>=4$. In Fig. \ref{fig:Incidence},
$s=1$ and $d=0.1067$ were used for the incidence curve based on $k_{min}=3$, 
while for $k_{min}=4$ we used $s=10$ and $d=0.028$. For the simulations using
$k_{min}=5$ and $k_{min}=6$ we set $s=7500$, $d=0.038$ 
and $s=10\,000\,000$, $d=0.0497$, respectively.

\section*{Results}

To create our model, we included empirical data representing age-related changes in neural
stem cell number and behavior. A population of neural stem cells is present in the human
brain at birth but declines exponentially thereafter \cite{Sanaietal2011corridors}.
Experiments in rodents demonstrate
that the exponential decline in neural stem cell number continues across the lifespan
\cite{Stolletal2011increased,Ahleniusetal2009neural}.
This depletion of the stem cell population is due to cell death and terminal differentiation.
We have therefore approximated the size of this cell population ($N(t)$) with an exponential interpolation
of the data from the human brain. Further experiments have demonstrated that the remaining
population of NSCs in the aged brain have dysregulated cell cycle kinetics \cite{Stolletal2011increased}.
Individual
remaining stem cells have an increased likelihood of re-entering the cell cycle, resulting
in an increased number of cell divisions in a given period of time ($r(t)$). We have approximated
this behavior using a linear interpolation. Our model incorporates these empirically-determined
changes in neural stem cell number and behavior (Fig. \ref{fig:mitotic_rate}).

NSCs accumulate mutations in every cell cycle. The process of genome replication during
cell division is imperfect, as a certain number of mutations occur and some of
these mutations will remain unrepaired. The number of mutations incurred during a single
cell division has been estimated \cite{FrankNowak2004problems}. According to their assessment,
we denote by $\mu=10e-7$ the probability for a gene in the coding region
to mutate due to a single cell division.
No single mutation leads to oncogenesis, so multiple
“hits” are necessary for complete oncogenic transformation \cite{Stolletal2013impact,GilPerotinetal2006loss}.
Cancer is characterized
by a number of cellular changes, including loss of cell cycle control, self-sufficiency in
growth factor signaling, resistance to anti-growth signals, escape from apoptosis, invasion and
neovascularization \cite{HanahanWeinberg2000hallmarks}. When Hanahan and Weinberg
first described these hallmarks of cancer,
they proposed that approximately six mutations would be required to dysregulate all six of
these cellular activities \cite{HanahanWeinberg2000hallmarks}. Yet now researchers appreciate
that mutation of a single
multi-functional protein can predispose alterations to multiple cellular activities
\cite{Mikheevetal2009syngeneic,Stranoetal2007mutant}.
Since the cell is dependent upon semi-redundant regulatory pathways to control cell cycle
progression and other activities \cite{Stolletal2013impact}, loss of one major tumor
suppressor is not sufficient to
create a tumor \cite{GilPerotinetal2006loss} and multiple regulators must be disrupted
to achieve oncogenic transformation
\cite{Mikheevetal2009syngeneic,Chowetal2011cooperativity}. Of the 18\,440 ($n_{total}$)
protein-encoding genes in the human, 522 have a causal role in human
cancer and $n_{glioma}=29$ of these (Table S1) have a demonstrated role in promoting
gliomagenesis \cite{Bamfordetal2004cosmic}. We assessed how many mutations in this set
of oncogenes are required to achieve tumor
formation. Based on this minimum number of mutations ($k_{min}$), our model 
computes the total probability for a single NSC to become oncogenically transformed.
This integrative probability is calculated by summing up the individual probabilities
according to the following equation:

\begin{eqnarray}
p(t) & = & \sum\limits_{i=k_{min}}^{29}{p_{i}(t),}
\label{eq:single_cancerogenic}
\end{eqnarray} where $p_{i}(t)$ denotes the probability for $i$ oncogenic mutations to have occurred at time
t. We have estimated $p_{i}(t)$ using the experimentally assessed parameters 
$N(t)$, $r(t)$ and $\mu$. Based on the number of protein-coding and
gliomagenesis-relevant genes, the probability for any one of the 29 oncogenes to 
become mutated from cell division is given by $p_{onc}=n_{glioma}\cdot \mu$.
Assuming that any gene mutates with equal probability, the occurrence of oncogenic mutations can be
approximated by the binomial distribution. It follows that $p_{i}(t)$ is given by:

\begin{eqnarray}
p_{i}(t) & = & \binom{R(t)}{i}\cdot p_{onc}^{i}\cdot (1-p_{onc})^{R(t)-i},
\label{eq:binom_form}
\end{eqnarray} where $R(t)$ is the number of cell divisions a NSC has
undergone until time $t$. It is computed by integrating the cell division rate $r(t)$ across
the age span until time t.

\begin{figure}[!ht]
  \centering
  \includegraphics[width=1.0\textwidth]{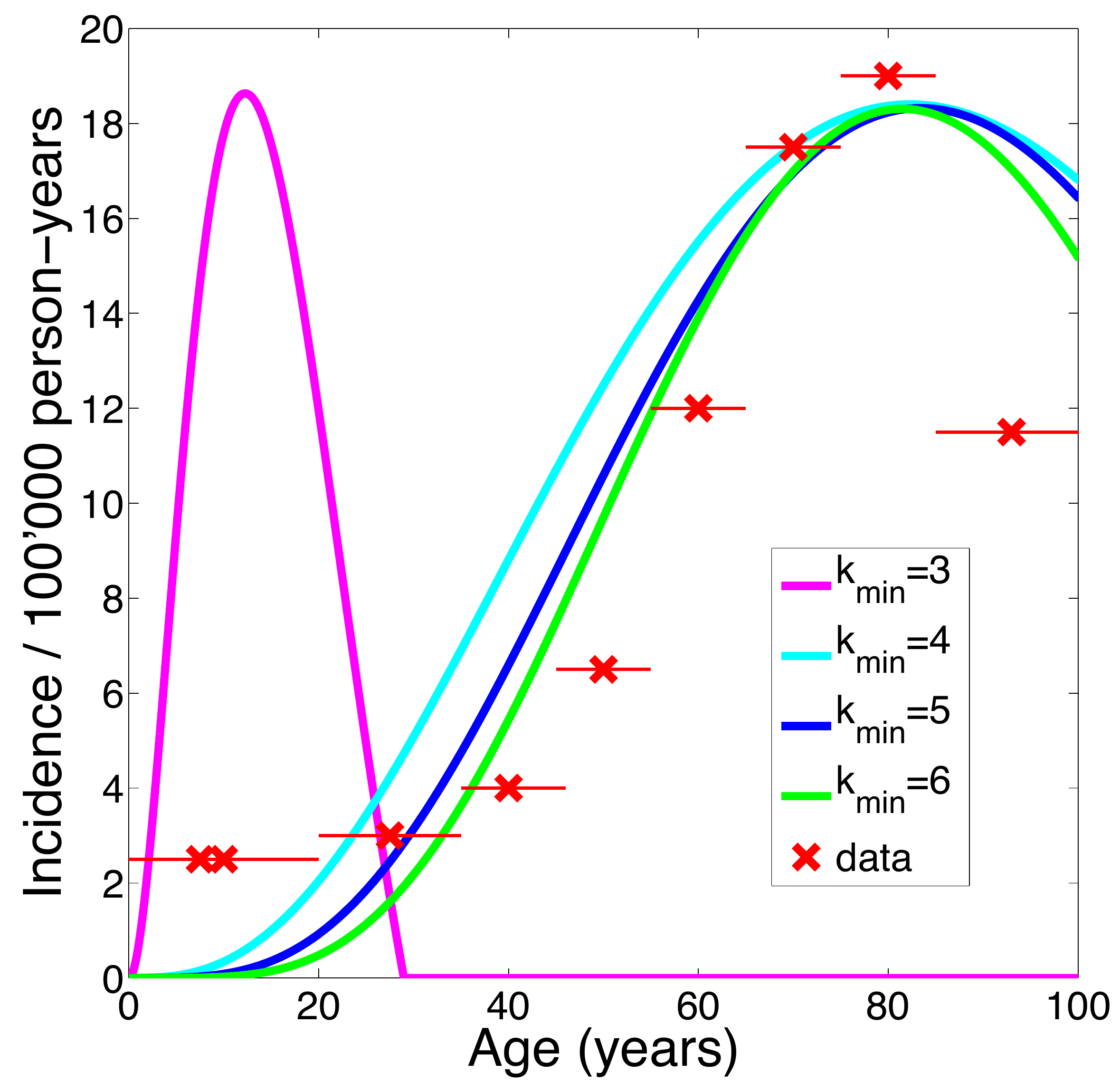}
\caption[Incidence]{{ \bf Influence of $k_{min}$ on location of peak incidence.}
Representative incidence curves for $k_{min}=3$ (magenta),
$k_{min}=4$ (cyan),
$k_{min}=5$ (blue) and $k_{min}=6$ (green). Only for
$k_{min}\geq 4$ can the condition of peak 
incidence at approximately 80 years be fulfilled. 
Incidence curves generated by the model for $k_{min}=4$, $5$ and $6$
are in accordance with the
demographic data from \cite{Doleceketal2012cbtrus} (red crosses: mean incidence 
of age groups, red lines: spans of age groups), with $k_{min}=6$ yielding the best
fit. Confidence intervals are shown in Suppl. Fig. S1. 
\label{fig:Incidence}}
\end{figure}

Given that $R(t)$ and $p_{onc}$ take sufficiently high ($>100$) and low ($<0.0001$)
values respectively, the Poisson
distribution is well-suited as an approximation for this otherwise computationally
very demanding formula:

\begin{eqnarray}
p_{i}(t) & = & \frac{\lambda^{i}}{i!}\cdot e^{-\lambda},
\label{eq:poissong_approx_form}
\end{eqnarray} with $\lambda(t)=R(t)\cdot p_{onc}$. The temporal sequence of 
oncogene mutations has been shown to be an important factor in tumor formation
\cite{Gerstungetal2011temporal,Guoetal2014inferring}, and so we have also
accounted for it in our 
model. Given that there are $i!$ possibilities for $i$ mutations to occur, Eq. \ref{eq:poissong_approx_form}
becomes:

\begin{eqnarray}
p_{i}(t) & = & \frac{s}{i!}\frac{\lambda^{i}}{i!}\cdot e^{-\lambda},
\label{eq:poissongapprox_ordered_form}
\end{eqnarray} where the scalar value $s$ represents the number of specific mutational
sequences necessary for oncogenic transformation. For $k_{min}=5$, we find
$s=7500$ to be an appropriate
value in order for the incidence curve to be in numerical accordance with the demographic
data (Fig. \ref{fig:Incidence}). This means that on average
$7500$ different sequences of mutations exist (for the different scenarios, i.e. 5, 6, ..., 29
oncogenes affected), which can ultimately lead to oncogenic transformation. 

The probability for a single cell to become oncogenically transformed is denoted by $p(t)$. 
Accordingly, the probability for glioma formation overall is proportional to the probability 
that at least one of all the NSC becomes transformed:

\begin{eqnarray}
  p_{glioma}(t)=1-(1-p(t))^{N(t)},
  \label{eq:minimum_onecellcancerogenic}
\end{eqnarray} where $N(t)=N_{0}e^{-d\cdot t}$ is the estimated number of NSCs
at time t. Hence, the parameter $d$ describes the decay of the NSC population over time, 
and so is in principle directly relatable to empirical data. We have
adapted $d$ such that the resulting incidence curve
matches the demographic data, while being qualitatively in accordance with experimental
findings in the mouse
\cite{Stolletal2011increased,Ahleniusetal2009neural}.

The prevalence of glioma is then proportional to $p_{glioma}(t)$. Since the 
units from the demographic datasets are with respect to $100\,000$ person-years, we 
compute the prevalence by multiplying $p_{glioma}(t)$ by $100\,000$.
From this, the incidence is computed 
by calculating the derivative. Since there are various 
time-varying parameters in the model, an analytical differentiation comprises a too
extensive formula. We therefore assess the incidence numerically. The obtained incidence curve
is shown in Fig. \ref{fig:Incidence} and resembles the demographic data.

The actual incidence of glioma across age demographics has been documented by The Central Brain
Tumor Registry of the United States \cite{Doleceketal2012cbtrus}.
We have used these published data to provide a
fit for the incidence and prevalence of glioma across the lifespan (Fig. \ref{fig:Incidence}).
The model parameters $s$ and $d$ were adapted in order to match with these incidence rates. The incidence
curves obtained from our model for $k_{min}=4$, $k_{min}=5$ or $k_{min}=6$ resemble
these demographic data. 
Also for $k_{min}>6$ is it possible to achieve agreement, and so our model 
yields a lower bound for the number of mutations required 
for oncogenic transformation. However, with increasing $k_{min}$ the model
parameters $s$ and $d$ need to change too. In particular, the parameter $s$ 
strongly increases. For $k_{min}=4$, $k_{min}=5$ and $k_{min}=6$ we find $s=10$, 
$s=7500$ and $s=10\,000\,000$ to be well-suited, respectively.

The biological
meaning of parameter $s$ in Eq. \ref{eq:poissongapprox_ordered_form} is twofold.
It captures that different oncogenes can
yield the same transformation 
hallmarks \cite{Hollandetal2000combined,Chowetal2011cooperativity},
and so multiple sequences of the same length could give rise
to glioma. Additionally, $s$ accounts for the possibility that 
different temporal sequences of the same oncogenes could lead to glioma 
formation. In 
the classical multistage model, there is only one temporal order that can achieve 
transformation. Importantly, since $s$ denotes an average number
of mutations, it could be different for different sequence lengths
$i$. With increasing $i$, $s_i$ can grow exponentially because of the factorials in the denominator 
of Eq. \ref{eq:poissongapprox_ordered_form}. For simplification and due to lack of detailed
empirical knowledge, we chose to use the same $s$ for all sequence lengths.

\begin{figure}[!ht]
    \centering
  \includegraphics[width=1.0\textwidth]{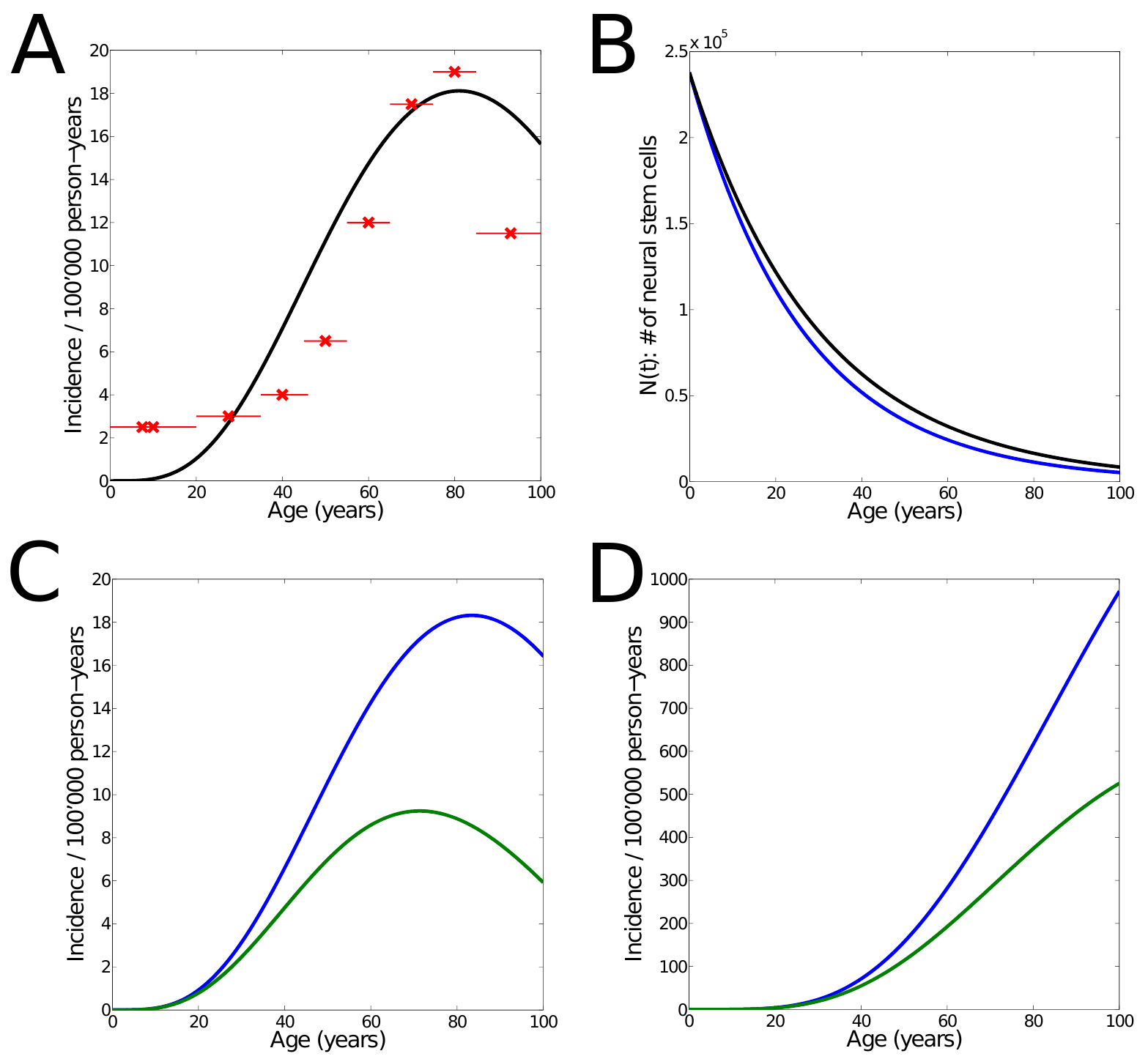}
\caption[Increasing \textit{vs.} constant cell division rate]{{\bf Effect of increasing
cell division rate.} {\bf (A)}
Modeled incidence of glioma (green) under constant cell division rate ($r(t)=251
\frac{divisions}{year}$). Model parameters $k_{min}=5$, $s=8800$ and
$d=0.0333$ were used in order to match with the demographic data (red crosses:
mean incidence of age groups, red lines: spans of age groups).
The increasing proliferation rate of NSCs is therefore not a necessary condition
for the 
incidence curve to match the demographic data, since similar results are obtained
after changes in the model parameters $s$ and $d$.
{\bf (B)} Number of NSCs over time, as used for the incidence curve shown in (A) (black) and
for the scenario where cell division rate increases linearly (Fig. \ref{fig:Incidence}, blue). Small
changes in the number of NSCs over time are sufficient to make up for the constant
cell division rate. It remains an empirical question which estimates of $N(t)$ and
$r(t)$ are correct in the adult human, since these are extrapolated from the model, the
young human, and the aging rodent. 
{\bf (C)} Incidence of glioma as derived from our model, for
increasing (blue) and constant (green) cell division rate during aging. Model parameters
are the same ($k_{min}=5$, $s=7500$, $d=0.038$). The green curve is the predicted
incidence by the model if the
proliferation rate was constant, and so leads an estimate of the net effect of the
increase. Overall,
our model suggests that the increase in cell-cycle re-entry substantially
increases glioma formation. {\bf (D)} 
Prevalence of glioma for increasing (blue) and constant (green) cell division 
rate. As shown in Suppl. Fig. S2,
the results are qualitatively confirmed also for $k_{min}$ = 4. 
\label{fig:incrVsconstantCellDivisionRate}}
\end{figure}

Since no studies in the human have directly demonstrated increased cell division in NSCs,
we have created a related model that assumes no age-related changes in cell division rate,
cell cycle length or likelihood to re-enter cell cycle. This adjusted model yields the
same results in glioma incidence and required mutation number if the exponential decrease
in proliferative cell number is adjusted accordingly (Fig. \ref{fig:incrVsconstantCellDivisionRate}A).
This age-related change is
therefore not a necessary condition of the model. Future labelling studies of the proliferative
cell population in the human brain will help to evaluate the relative accuracy of these
two models. Interestingly, the model quantifies the net effect of an increasing cell
division rate while the other parameters are the same
(Fig. \ref{fig:incrVsconstantCellDivisionRate}CD). These results suggest that
this increase of cell division rate almost doubles the
occurrence of glioma.

\section*{Discussion}

Mathematical modeling has been used to create predictions regarding the growth of tumors
\cite{Andersonetal2006tumor,Choeetal2011model} and response of individual tumors to
surgical resection or radiotherapy \cite{Swansonetal2008mathematical,Rockneetal2010predicting}.
The incidence of tumors in a human population has also been modeled
\cite{ArmitageDoll1954age,PompeiWilson2002quantitative}. However these
models of cancer incidence did not employ empirical measures of age-related changes in cellular
dynamics, nor did they incorporate experimental knowledge on 
glioma-related proto-oncogenes. Here we present a model to predict the probability of glioma incidence across the lifespan
based on neural stem cell dynamics in the individual organism. 

We find that a simple model using recent estimates of biological parameters on the single-cell
level can account for demographic observations. Along these lines, we provide a modified and
extended version of the well-established Armitage-Doll model \cite{ArmitageDoll1954age}.
In contrast to this
classical approach, we do not restrict our model to a specific number of oncogenic mutations.
Instead, we account for all the numbers of oncogenic mutations that possibly can occur
(i.e. mutations of $k_{min}$ to 29 oncogenes, see Eq. \ref{eq:single_cancerogenic}). Our model
therefore does not rely on the
(experimentally unsupported) assumption of the classical Armitage-Doll model that
only a specific number of oncogenes must be mutated for oncogenic transformation.

Since the parameters of our model have a direct biological meaning,
further biological data can be incorporated and predictions can be made.
For example, previous theories have yielded various estimates for the minimal number of
oncogenic mutations required for carcinogenesis \cite{ArmitageDoll1957two,Fisher1958multiple,ArmitageDoll1961stochastic}.
Notably, we come to the conclusion that a minimum of 4 or 5 oncogenic mutations is sufficient for 
tumorigenesis, in contrast to 6-7 mutations as implicated by the classical Armitage-Doll model
\cite{ArmitageDoll1954age} and as predicted by Hanahan and Weinberg \cite{HanahanWeinberg2000hallmarks}.
$k_{min}=5$ is higher than experimental results which demonstrate that NSCs can be oncogenically
transformed successfully with only three oncogenic mutations specifically affecting the PTEN,
p53 and Rb pathways \cite{Mikheevetal2009syngeneic,Chowetal2011cooperativity,FriedmannMorvinskietal2012dedifferentiation}.
However, many human gliomas regardless of grade demonstrate 5 mutations, namely affecting
EGFR, PTEN, $P16_{INK4A}$, TP53 and MDM2 \cite{OstromBarnholtzSloan2011current}.
Therefore our model is in line with empirical studies on the number
of mutations required to achieve oncogenic transformation. Many mutations affecting tumor suppressor
pathways will cause a cell to undergo senescence,
slowing the cell division rate and increasing the likelihood of apoptosis.
Very few sequences of mutation are likely to bypass this protective response.
So it is easy to imagine that few scenarios ($s=10$) are compatible with a low number
of mutations achieving oncogenic transformation ($k_{min}=4$), while more scenarios
($s=7500$) can achieve oncogenic transformation with a larger number of mutations ($k_{min}=5$).
Considering that different oncogenic mutations yield the same hallmark, and that 
multiple temporal sequences of the same mutations could yield the same result, we find
$s=7500$ more plausible than $s=10$. This model therefore supports the conclusion that five
oncogenic mutations are sufficient to achieve oncogenic transformation and
initiate gliomagenesis.

Our model accounts for
the possibility that some oncogenes, due to more
interactions, play a more central role than others \cite{Jeongetal2001lethality}. Therefore, fewer
mutations of such hub genes might be sufficient for the formation of glioma.
It is possible that altered function of such hub genes could lead to genomic instability and
increased mutation rate. However, one assumption in our model is the stable
accumulation of mutations in every cell cycle. While
this number of mutations have been estimated in proliferative cell types \cite{FrankNowak2004problems},
this rate may indeed depend on prior changes. With age, the genome becomes more unstable due to shortened
telomeres, increased mutation load and chromosomal abnormalities \cite{Baileyetal2004accumulation}.
All of these changes
could increase the likelihood of mutations or disrupt the efficacy of repair mechanisms. The
net mutations incurred during each division may therefore increase with age. However any
age-related changes to the mutation rate depending on prior mutation load have not been
empirically determined so we were unable to incorporate this age-related factor into our
calculations. We have therefore estimated that the mutation rate remains constant across
the lifespan.

However our model does allow us to incorporate different weightage for mutations, i.e. that some
mutations are less likely to co-exist than others, as has been established by the Cancer
Genome Atlas effort (\texttt{http://cancergenome.nih.gov/})
\cite{GilPerotinetal2006loss,Gerstungetal2011temporal}.
In Eq. \ref{eq:poissongapprox_ordered_form}
the denominator increases much faster than the
nominator with the length of the modeled sequence of oncogenes, and so long sequences are
unlikely to occur. Hence, mutational combinations that are included only in the long sequences
are unlikely to co-exist overall.

In light of evidence that a temporal sequence of mutations may be
crucial in tumorigenesis \cite{Gerstungetal2011temporal, Guoetal2014inferring}, it is
notable that our model considers variation in the number and order of oncogenic mutations
needed to invoke glioma formation. 
Our model thus usefully explores the relationship between
these experimentally tractable variables, particularly
$k_{min}$, $d$, $s$, $N(t)$ and $r(t)$.

Similar to previous researchers \cite{PompeiWilson2002quantitative}, we have included
an age-related decline in the number
of proliferative cells, which is responsible for the characteristic peak of the incidence
at 80 years. In contrast to their linear decrease, we model an exponential decrease of the
proliferative pool which matches better with experimental findings in this cell population
\cite{Stolletal2011increased,Sanaietal2011corridors,Ahleniusetal2009neural}.
In addition, we employ empirically-derived results to estimate cell cycle
length [12], the mutation rate during each cell cycle \cite{FrankNowak2004problems}
and the fraction of genes that
promote oncogenic transformation upon mutation (\cite{Bamfordetal2004cosmic}
and Table S1). Together, these data can
be used to predict the age-associated incidence of glioma in the human population 
\cite{Doleceketal2012cbtrus}
without the need to assert differential susceptibility throughout the population which is
not supported by biological evidence \cite{Ritteretal2003multistage}.

It is possible that other cell types besides the neural stem cell give rise
to glioma. One recent study demonstrated that mature cells such as neurons can be forced
to undergo oncogenic transformation using cell-specific targeting of two major tumor suppressor
pathways \cite{FriedmannMorvinskietal2012dedifferentiation}, however it is not clear
that such mutations could randomly occur in a post-mitotic
cell population. Alternatively, glial progenitor cells within the white matter have been
proposed to be the true glioma cell-of-origin \cite{Liuetal2011mosaic,Assanahetal2006glial}.
Empirical data on these cells are
scarcer, so we are currently unable to estimate the size of this population and the rate of
glial progenitor cell division across the lifespan (key variables for implementing this model).
Future studies may help to address whether the cell cycle kinetics of this population can also
predict actual glioma incidence in the human population. Variability in the cellular origin
as well as the underlying genetic lesions of glioma could in part explain the extraordinary
heterogeneity in this tumor type. Yet the evidence most strongly implicates the multi-potent
neural stem cell as the most likely cell of origin, so we have focused on this cell type
in our model.

There is evidence to suggest the molecular pathogenesis of high-grade gliomas
(presenting as primary glioblastoma) is different to that of low-grade gliomas
(presenting as grade II-III astrocytoma or oligodendroglioma, often progressing to
secondary glioblastoma). These two types of brain tumor have different genetic and
epigenetic profiles, with different initiating mutations \cite{OhgakiKleihues2013definition}.
In the future, this
model could be adapted to include such different constraints on molecular pathogenesis
to distinguish between the incidence rates of low-grade and high-grade glioma.

Overall, we provide a model that uses experimentally obtained parameters on neural stem
cell proliferation and yields results which match with actual demographic data in the human
population. We demonstrate the consistency of our model which incorporates estimates of
neural stem cell number, cell division rate, mutation rate and number of oncogenes. Importantly,
our model supports the hypothesis that glioma is caused by randomly occurring oncogenic mutations
within the neural stem cell population of the adult brain.

\section*{Acknowledgments}

This work is dedicated to Thomas L{\"o}tsch, who deceased after his combat against cancer.
His inspirational spirit and zest for life will prevail.

We thank Peter Taylor, Yujiang Wang and Fr\'ed\'eric Zubler for helpful suggestions to the manuscript.

\section*{Funding}

R. B. and M. K. were supported by the Human Green Brain Project (http://www.greenbrainproject.org)
through the Engineering and Physical Sciences Research Council (EP/K026992/1). The funders
had no role in study design, data collection and analysis, decision to publish, or
preparation of the manuscript.

\bibliography{biblio}
\newpage

\section*{Figures}

\newpage

\renewcommand{\thefigure}{S\arabic{figure}}
\setcounter{figure}{0}

\clearpage

\section*{Supplementary Information}

\begin{figure}[h!]
\centering
\includegraphics[width=1.0\textwidth]{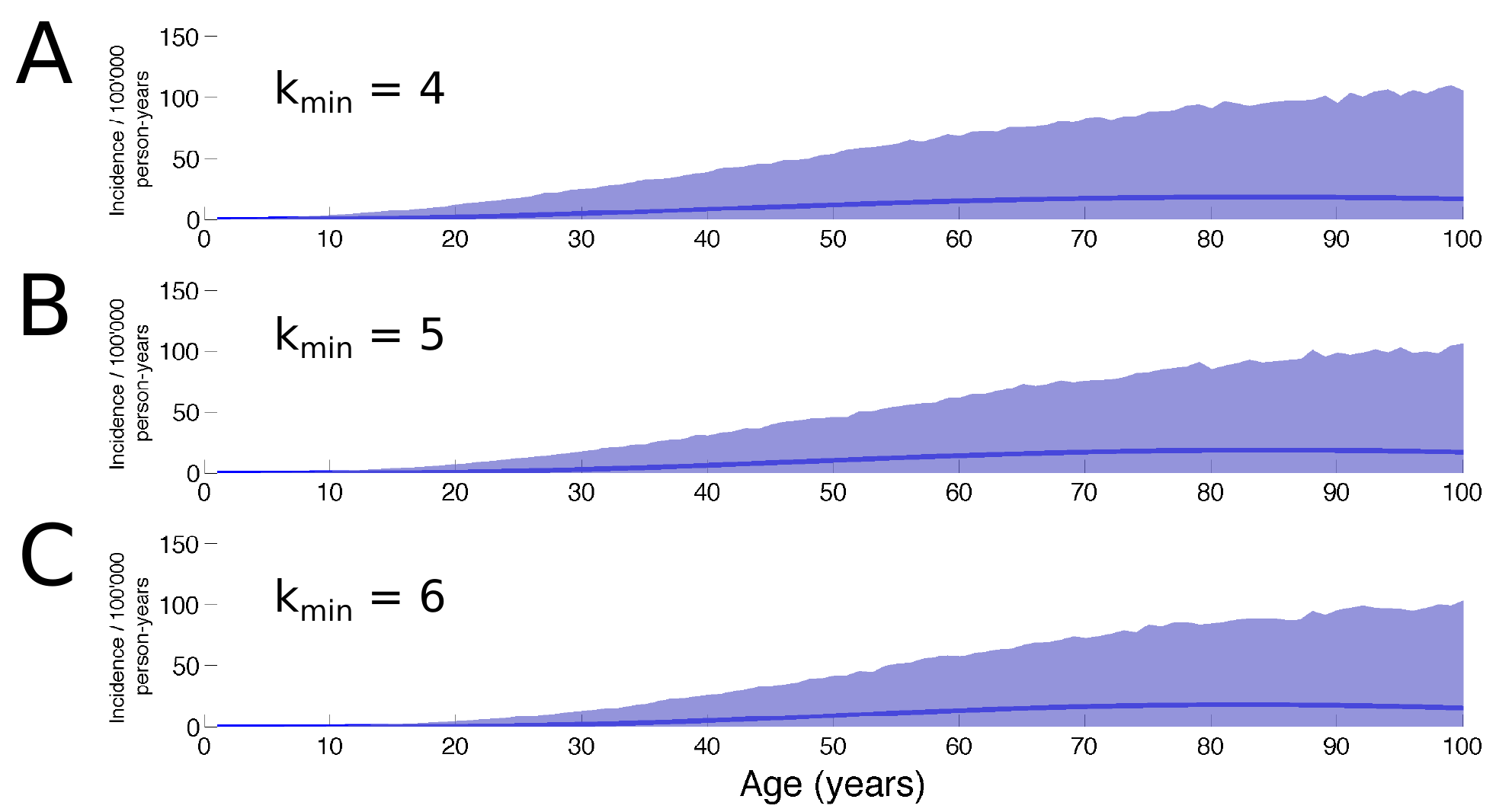}
\caption[CI incidence]{{\bf Confidence intervals
for modeled incidence.} 95 $\%$ confidence intervals (shaded) for the modeled incidence rates 
during aging, as computed by bootstrapping. The modeled incidence curve (blue line) 
is the same as shown in Fig. \ref{fig:Incidence} using
{\bf (A)} $k_{min}=4$, $s=10$ and $d=0.028$, {\bf (B)} $k_{min}=5$, $s=7500$ and $d=0.038$
and {\bf (C)} $k_{min}=6$, $s=10\,000\,000$ and $d=0.0497$.
\label{fig:CI_incidence}}
\end{figure}

\begin{figure}[h!]
\centering
\includegraphics[width=1.0\textwidth]{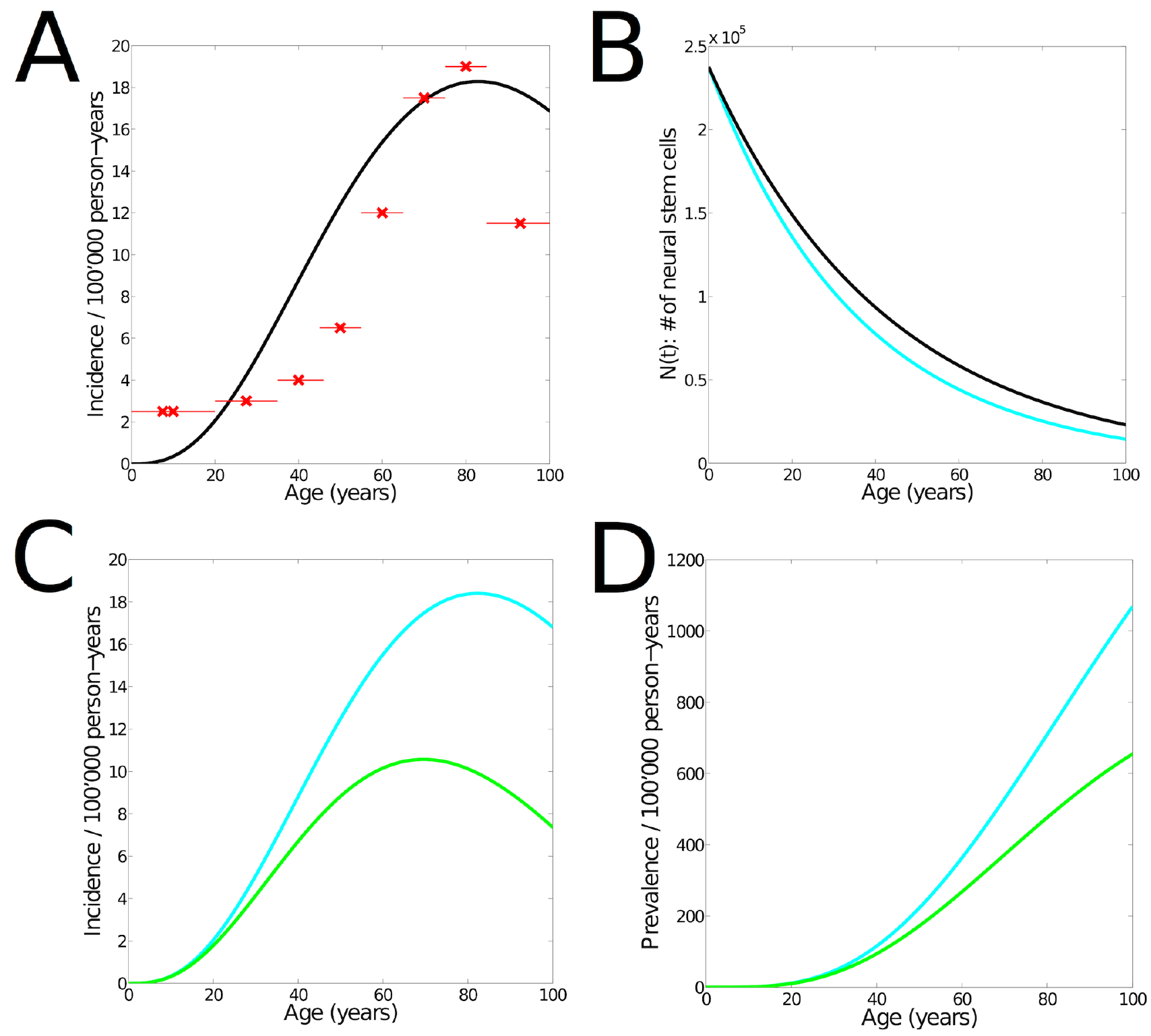}
\caption[Increasing \textit{vs.} constant cell division rate for $k_{min}=4$]
{{\bf Effect of increasing
cell division rate for scenario with $\mathbf{k_{min}=4}$.} {\bf (A)}
Modeled incidence of glioma (black) under constant cell division rate ($r(t)=251
\frac{divisions}{year}$). Model parameters $k_{min}=4$, $s=10.2$ and
$d=0.0233$ were used in order to match with the demographic data (red crosses:
mean incidence of age groups, red lines: spans of age groups).
The increasing proliferation rate of NSCs is therefore not a necessary condition
for the 
incidence curve to match the demographic data, since similar results are obtained
after changes in the model parameters $s$ and $d$.
{\bf (B)} Number of NSCs over time, as used for the incidence curve shown in (A) (black) and
for the scenario where cell division rate increases linearly (Fig. \ref{fig:Incidence}, cyan).
Small changes in the number of NSCs over time are sufficient to make up for the constant
cell division rate. It remains an empirical question which estimates of $N(t)$ and
$r(t)$ are correct in the adult human, since these are extrapolated from the model, the
young human, and the aging rodent. 
{\bf (C)} Incidence of glioma as derived from our model, for
increasing (cyan) and constant (green) cell division rate during aging. Model parameters
are the same ($k_{min}=4$, $s=10$, $d=0.028$). The green curve is the predicted
incidence by the model if the
proliferation rate was constant, and so leads an estimate of the net effect of the
increase. Overall, as for $k_{min}=5$
our model suggests that the increase in cell-cycle re-entry substantially
increases glioma formation. {\bf (D)} 
Prevalence of glioma for increasing (cyan) and constant (green) cell division 
rate.
\label{fig:incrVsConstMitratekmin6}}
\end{figure}


\clearpage
\renewcommand{\thetable}{S\arabic{table}}
\newpage

\begin{table}[!ht]
\caption*{
\bf{Table S1: Proto-oncogenes implicated in glioma formation}}
\hspace{-2 cm} 
\begin{tabular}{|c|p{4cm}|c|c|c|p{5cm}|}
\hline
\textbf{Symbol} & \textbf{Name} & \textbf{GeneID} & \textbf{Chromosome} & \textbf{Chr Band} & \textbf{Tumor Types (Somatic Mutations)}
\\ \hline
APC & adenomatous polyposis of the colon gene & 324 & 5 & 5q21 & colorectal, pancreatic, desmoid, hepatoblastoma, glioma, other CNS
\\ \hline
BRAF	 & v-raf murine sarcoma viral oncogene homolog B1 & 673 & 7 & 7q34 & melanoma, colorectal, papillary thyroid, borderline ovarian, NSCLC, cholangiocarcinoma, pilocytic astrocytoma 
\\ \hline
CDKN2A & cyclin-dependent kinase inhibitor 2A (p16(INK4a)) gene & 1029 & 9 & 9p21 & melanoma, multiple other tumor types 
\\ \hline
CDKN2a(p14) & cyclin-dependent kinase inhibitor 2A--  p14ARF protein & 1029 & 9 & 9p21 & melanoma, multiple other tumor types
\\ \hline
CDKN2C & cyclin-dependent kinase inhibitor 2C (p18, inhibits CDK4) & 1031 & 1 & 1p32 
& glioma, MM
\\ \hline
CIC & capicua homolog & 23152 & 19 & 19q13.2 & oligodendroglioma, soft tissue sarcoma
\\ \hline
COPEB & core promoter element binding protein (KLF6) & 1316 & 10 & 10p15 & prostate, glioma
\\ \hline
CTNNB1 & catenin (cadherin-associated protein), beta 1 & 1499 & 3 & 3p22-p21.3 & 
colorectal, ovarian,  hepatoblastoma, pleomorphic salivary gland adenoma, other tumor types
\\ \hline
EGFR & epidermal growth factor receptor & 1956 & 7 & 7p12.3-p12.1 & glioma, NSCLC
\\ \hline
ERBB2 & v-erb-b2 erythroblastic leukemia viral oncogene homolog 2 & 2064 & 17 & 17q21.1 
& breast, ovarian, other tumor types, NSCLC, gastric
\\ \hline
FUBP1 & far upstream element (FUSE) binding protein 1 & 8880 & 1 & 1p13.1 & oligodendroglioma
\\ \hline
GOPC & golgi associated PDZ and coiled-coil motif containing & 57120 & 6 & 6q21 
& glioblastoma
\\ \hline
H3F3A & H3 histone, family 3A & 3020 & 1 & 1q42.12 & glioma
\\ \hline
HIST1H3B & histone cluster 1, H3b & 3020 & 6 & 6p22.1 & glioma
\\ \hline
IDH1 & isocitrate dehydrogenase 1 (NADP+), soluble & 3417 & 2 & 2q33.3 & glioblastoma
\\ \hline 
\end{tabular}
\label{tab:label}
\end{table}

\begin{table}[!ht]
\caption*{
\bf{Table S1 (continued). Proto-oncogenes implicated in glioma formation}}
\hspace{-2 cm} 
\begin{tabular}{|c|p{4cm}|c|c|c|p{5cm}|}
\hline
\textbf{Symbol} & \textbf{Name} & \textbf{GeneID} & \textbf{Chromosome} & \textbf{Chr Band} & \textbf{Tumor Types (Somatic Mutations)}
\\ \hline
IDH2 & socitrate dehydrogenase 2 (NADP+), mitochondrial & 3418 & 15 & 15q26.1 & glioblastoma
\\ \hline
KIAA1549 & KIAA1549 & 57670 & 7 & 7q34 & pilocytic astrocytoma
\\ \hline
KRAS & v-Ki-ras2 Kirsten rat sarcoma 2 viral oncogene homolog & 3845 & 12 & 12p12.1 
& pancreatic, colorectal, lung, thyroid, AML, other tumor types
\\ \hline
MDM2 & Mdm2 p53 binding protein homolog & 4193 & 12 & 12q15 & sarcoma, glioma, colorectal, other tumor types
\\ \hline
MDM4 & Mdm4 p53 binding protein homolog & 4194 & 1 & 1q32 & glioblastoma, bladder, retinoblastoma
\\ \hline
MYC & v-myc myelocytomatosis viral oncogene homolog (avian) & 4609 & 8 & 8q24.12-q24.13  
& Burkitt lymphoma, amplified in other cancers, B-CLL
\\ \hline
NF1 & neurofibromatosis type 1 gene & 4763 & 17 & 17q12 & neurofibroma, glioma
\\ \hline
PIK3CA & phosphoinositide-3-kinase, catalytic, alpha polypeptide & 5290 & 3 & 3q26.3 
& colorectal, gastric, glioblastoma, breast
\\ \hline
PIK3R1 & phosphoinositide-3-kinase, regulatory subunit 1 (alpha) & 5295 & 5 & 5q13.1 
& glioblastoma, ovarian, colorectal
\\ \hline
PTEN & phosphatase and tensin homolog gene & 5728 & 10 & 10q23.3 & glioma, prostate, endometrial
\\ \hline
RAF1 & v-raf-1 murine leukemia viral oncogene homolog 1 & 5894 & 3 & 3p25 & pilocytic astrocytoma
\\ \hline
ROS1 & v-ros UR2 sarcoma virus oncogene homolog 1 (avian) & 6098 & 6 & 6q22 & glioblastoma, NSCLC
\\ \hline
SRGAP3 & SLIT-ROBO Rho GTPase activating protein 3 & 9901 & 3 & 3p25.3 & pilocytic astrocytoma
\\ \hline
TP53 & tumor protein p53 & 7157 & 17 & 17p13 & breast, colorectal, lung, sarcoma, adrenocortical, glioma, multiple other tumor types
\\ \hline
\end{tabular}
\caption*{
{\bf Table S1: Proto-oncogenes implicated in glioma formation.} Information on the 29 proto-oncogenes that have been implicated in the formation of glioma. The COSMIC Cancer Gene Census is a regularly-updated catalogue of somatic cell
mutations causally implicated in cancer: \texttt{http://cancer.sanger.ac.uk/cosmic/census}.
Of all genes listed, we have selected genes with a known role in glioma (including subtypes such as
glioblastoma, astrocytoma, oligodendroglioma). An additional 6 genes were listed in the
COSMIC gene database as being implicated in {}``other tumor types''. These genes, KRAS,
MYC, CDKN2A(p16), CDKN2A(p14),
CTNNB1(beta-catenin), and ERBB2(HER2), have indeed been implicated in gliomagenesis
in other studies \cite{Chietal2012prospective, Guanetal2014molecular, Waageetal2013c, Lietal2014expression},
so we have included them in this list.
The probability of any one
of the oncogenes being mutated is equivalent to 
$p_{onc}=n_{glioma}\cdot\mu$, where $n_{glioma}$ is the number of
oncogenes involved in glioma formation and $\mu$ is the probability for genetic mutation
due to a single cell division.
\label{tab:label}}
\end{table}

\end{document}